\documentclass[useAMS,usenatbib,usegraphicx,dvips]{mn2e}

\title[Sources of Radio Background]{Sources of the Radio Background Considered}
\author[J.~Singal et al.]{J. Singal$^{1}$\thanks{E-mail: \texttt{jsingal@stanford.edu}}, {\L}.~Stawarz$^{1,\,2,\,3}$, A.~Lawrence$^{4}$\thanks{Visiting physicist at KIPAC/SLAC National Accelerator Laboratory and Stanford University, USA}, V.~Petrosian$^{1}$\thanks{Also Departments of Physics and Applied Physics, Stanford University, Stanford, CA 94305, USA}\\ 
$^{1}$Kavli Institute for Particle Astrophysics and Cosmology, SLAC National Accelerator Laboratory and Stanford University,\\
382 Via Pueblo Mall, Stanford, CA 94305-4060, USA\\ 
$^{2}$Institute of Space and Astronautical Science, Japan Aerospace Exploration Agency,\\ 
3-1-1 Yoshinodai, Chuo-ku, Sagamihara, Kanagawa 252-5510, Japan\\ 
$^{3}$Astronomical Observatory of the Jagiellonian University, ul. Orla 171, 30-244 Krak\'ow, Poland\\
$^{4}$University of Edinburgh Institute for Astronomy, Scottish Universities Physics Alliance, Royal Observatory,\\ 
Blackford Hill, Edinburgh UK} 

\begin{document}

\date{In press version, 2010 August 16}

\pagerange{\pageref{firstpage}--\pageref{lastpage}} \pubyear{2010}

\maketitle

\label{firstpage}

\begin{abstract}
We investigate different scenarios for the origin of the extragalactic radio background.  The surface brightness of the background, as reported by the ARCADE 2 collaboration, is several times higher than that which would result from currently observed radio sources.  We consider contributions to the background from diffuse synchrotron emission from clusters and the intergalactic medium, previously unrecognized flux from low surface brightness regions of radio sources, and faint point sources below the flux limit of existing surveys.  By examining radio source counts available in the literature, we conclude that most of the radio background is produced by radio point sources that dominate at sub $\mu$Jy fluxes.  We show that a truly diffuse background produced by electrons far from galaxes is ruled out because such energetic electrons would overproduce the obserevd X-ray/$\gamma$-ray background through inverse Compton scattering of the other photon fields.  Unrecognized flux from low surface brightness regions of extended radio sources, or moderate flux sources missed entirely by radio source count surveys, cannot explain the bulk of the observed background, but may contribute as much as 10 per cent.  We consider both radio supernovae and radio quiet quasars as candidate sources for the background, and show that both fail to produce it at the observed level because of insufficient number of objects and total flux, although radio quiet quasars contribute at the level of at least a few percent.  We conclude that if the radio background is at the level reported, a majority of the total surface brightness would have to be produced by ordinary starforming galaxies above redshift 1 characterized by an evolving radio far-infrared correlation, which changes toward the radio loud with redshift.
\end{abstract}


\section{Introduction}

Recent results from the ARCADE 2 (Absolute Radiometer for Cosmology, Astrophysics, and Diffuse Emission) project suggest a radio background several times brighter than can be produced by currently observed radio sources \citep{Fixsen09,Seiffert09,Singal09}.  The extragalactic background, detected from 22 MHz to 8 GHz, has a power law spectral index of $\alpha \simeq 0.6$ (defined here as $S_{\nu} \propto \nu^{-\alpha}$, where $S_{\nu}$ is the spectral flux density) and a brightness temperature of $1.17$\,K at $1$\,GHz, corresponding to a cosmic radio background (CRB) intensity of $7\times 10^{-22}$ W Hz$^{-1}$ m$^{-2}$ sr$^{-1}=3.6\times 10^4$ Jy/sr.  A significant contribution to the background from free-free emission has been ruled out based on the spectral shape \citep{Seiffert09}.

It is unlikely that the radio signal reported by the ARCADE 2 collaboration is Galactic or local in origin.  As discussed in \citet{Kogut09}, the extragalactic component is separated from the diffuse Galactic foreground by two independent robust indicators, a correlation of radio with CII emission and a cosecant dependence on Galactic latitude.  The inferred residual extragalactic component is several times brighter than the high latitude Galactic level.  Comparisons with observed edge-on galaxies indicate that our Galaxy would be quite anomolous to support a radio emitting halo of the intensity needed to explain the signal\footnote{One may add here that this is true even taking into account presence of `anomalous large-scale radio continuum features' present in some edge-on spirals \citep{hum83,elm95}.}.  Additionally, assuming a diffuse Galactic origin for the measured radio signal would require a source of radio emission that does not follow the correlation with far-infrared emission observed in local galaxies, again making our Galaxy anomalous.  Isotropic radio emission from a local region on the scale of the local bubble accounting for the level observed would manifest an all-sky quadrupole polarization pattern at a level visible in WMAP 23 GHz data, but such a pattern is not seen.  These lines of evidence are summarized in \citet{Kogut09}.  Furthermore, the considerations of \S \ref{xrb} in this paper can be used to place a strong constraint on the amount of the total observed high latitude radio emission that can Galactic in origin, as in the halo inverse compton scattering on the ambient light would produce a much stronger X-ray background than observed, again strongly disfavoring a Galactic origin as an explaination for the signal.   In this paper, then, we investigate possible origins for the radio background, assuming the extragalactic level is that reported by the ARCADE 2 collaboration, although we conclude that producing the background at the level observed is difficult without a mechanism to cause significant redshift evolution in the radio to far infrared correlation beyond what has been observed thus far.  

Interferometric surveys have achieved a relatively consistent picture of radio source counts above the $10 \mu$Jy level. At high fluxes, say between $1$\,mJy and $1$\,Jy at 1.4 GHz, radio loud active galactic nuclei (AGN) dominate radio source counts \citep[e.g.,][]{Condon07,Windhorst93}, with their differential counts following a power law $dN/dS\propto S^{-\gamma}$ with $\gamma<2.5$ (the so-called Euclidean value). Below $1$\,mJy, this trend tends to reverse and the source counts show a more rapid increase with decreasing flux, perhaps indicating emergence of a new population which does not contribute at higher fluxes but dominates in this regime \citep[e.g.,][]{Condon07,Simpson06,Hopkins03}. \citet{Gervasi08} have calculated the radio brightness resulting from fitting a two population model and extrapolating available radio source counts to fainter fluxes at frequencies from $150$\,MHz to $8.5$\,GHz. They derive a surface brightness that is 3 to 6 times smaller than that reported by the ARCADE 2 collaboration. Clearly there must be significant contributions to the radio background from sources not considered by \citet{Gervasi08}.

ARCADE 2 and other instruments used to determine the diffuse background intensity have resolutions greater than 1 degree, and cannot distinguish between a background due to discrete sources of angular sizes smaller than one degree, and ones that are truly diffuse. Therefore, we consider several candidates for comprising the background.

The emission mechanism of sources producing the CRB must be synchrotron radiation by relativistic electrons. This puts powerful constraints on potential sources.  They must not overproduce the measured diffuse far-infrared background if related to star forming activity, nor overproduce the measured diffuse X-ray/$\gamma$-ray background through inverse compton (IC) scattering of the cosmic microwave background (CMB) and other background photon fields by the same electrons.  Additionally, the source spectra must be consistent with the CRB power-law spectrum of index $0.6$, mentioned above.

In this paper, we investigate several kinds of possible candidates. In \S\,\ref{sourcecounts} we analyze the existing data from radio source counts, and using some approximate fitting procedures estimate the contribution from sources observed in numerous surveys. In \S\,\ref{xrb} we use the X-ray/$\gamma$-ray background to limit the contribution from more diffuse emission, clusters, and the intergalactic medium, and in \S\,\ref{mlsb} we examine whether source count surveys have missed a significant amount of flux.  In \S\,\ref{rsn} and \S\,\ref{rq} we explore the potential contribution from radio supernovae and radio quiet quasars, and finally, in \S\,\ref{gals} we conclude that emission from star forming galaxies may account for bulk of the measured radio background, but only if the radio/far-infrared luminosity ratio increases with redshift.  A brief summary and discussion is presented in \S\, \ref{discussion}.

\section{Radio Source Counts} \label{sourcecounts}

\begin{figure}
\includegraphics[width=3.5in]{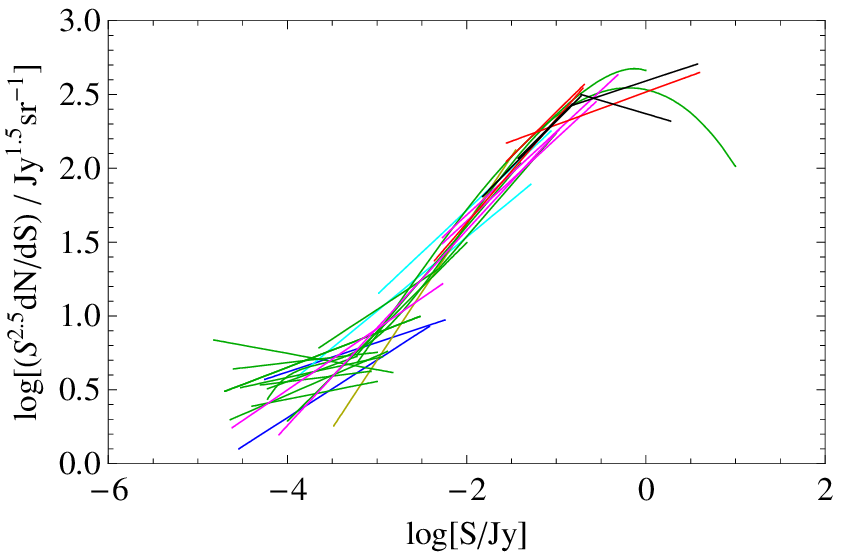}
\includegraphics[width=3.5in]{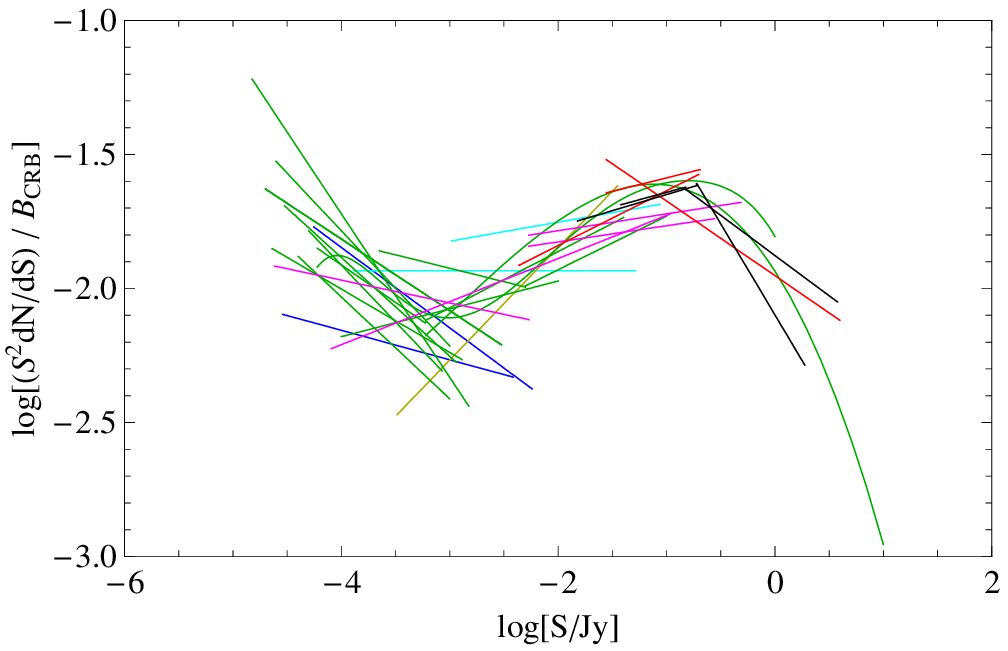}
\caption{{\bf TOP:} Scheamatic description of the observed $S^{2.5}dN/dS$ distribution from  different radio surveys. Source counts from $8.4$\,GHz surveys are in dark blue, $2.7$\,GHz in light blue, $5$\,GHz in yellow, $1.4$\,GHz in green, $610$\,MHz in pink, $408$\,MHz in red, and $151$\,MHz in black. Details of the radio surveys used with references are given in Table\,\ref{sctable}. Data from all frequencies is normalized to flux bins at $1.4$\,GHz using a $0.75$ spectral index.
{\bf BOTTOM:} The $S^2dN/dS$ distribution representing the  surface brightness per logarithmic flux bin of the  same surveys, divided by the surface brightness of CRB at 1.4 GHz reported by the ARCADE collaboration. The source counts tell roughly a consistent story over a wide range of frequencies and resolutions.  The peak icontribution at $\sim 100$\,mJy comes from bright radio galaxies. The contribution  decreases with decreasing flux density, but begins to rise again below $\sim 1$\,mJy as a new population of sources becomes important. Integrating over the entire flux range probed by these surveys gives $\sim 26$ per cent of the total surface brightness of the CRB at $1.4$ GHz.}
\label{scbfigure}
\end{figure}

Figure\,\ref{scbfigure} shows a schematic but fairly accurate depiction of radio source counts available in the literature (see Table\,\ref{sctable}). In the top panel of the figure, we plot the $S^{2.5}dN/dS$ distributions given in different radio surveys (note that the differential count is per steradian per unit flux).  The observations were conducted at a range of frequencies ($0.151-8.5$\,GHz). We have converted all  fluxes to $1.4$\,GHz assuming  a general power-law spectrum with an average (redshift independent) spectral index equal to $0.75$, the canonical value of the extragalactic radio sources.  Note that even though the surveys considered have a wide range of angular resolution (spanning from $1$ to $300$ arcseconds) there is not only a clear agreement among the counts at  $S > 1$\,mJy, but  also in regards to the presence of the -- relatively poorly constrained -- low-flux population emerging at  $S < 1$\,mJy.

In the bottom panel of Figure\,\ref{scbfigure} we plot a proxy for 1.4 GHz total surface brightness due to radio sources,   $S^2 \, (dN/dS)$, divided by the observed CRB brightness as reported by the ARCADE collaboration in \citet{Fixsen09}, i.e.
\begin{equation}
\left. {S^2 \, (dN/dS) \over B_{CRB}\!(\nu)}\right|_{\rm 1.4\,GHz} \equiv \left. {S^2 \, (dN/dS) \over 2 k \nu^2 c^{-2} \,T_{CRB}\!(\nu)}\right|_{\rm 1.4\,GHz} \, , 
\end{equation}

\noindent
where
\begin{equation}
 T_{CRB}\!(\nu) = 1.17 \times \left({\nu \over {\rm GHz}}\right)^{-2.6} \, {\rm K} \, . 
\end{equation}
\noindent
The true surface brightness 
\begin{equation}
B_{counts}(S) = \int^\infty_s S \, {dN \over dS} \, dS  =  {1 \over (\gamma - 2)} \, S^2 \, {dN \over dS} \,
\end{equation}
for a power law $dN/dS \equiv S^{-\gamma}$, is equal to the area under the $S^2 (dN/dS)$ curve, and thus the bottom pannel of Figure \ref{scbfigure} provides directly a minimum fractional contribution to the CRB from the resolved objects. We estimate this fraction to be $\simeq 26$ per cent. We note that the high-flux population is dominated by bright radio-loud AGN, while the low-flux one is generally thought to be dominated by starforming galaxies \citep[see][and references therein]{bal09}, with an additional possible contribution from radio-quiet and low-radio power AGN \citep[e.g.][]{Ibar09,pad09}.  Surveys probing fluxes as low as a few $\mu$Jy are consistent with the two population model \citep[e.g.,][]{Fomalont02,BI06}.

\subsection{High-Flux Population} \label{hf}

The integrated (fractional) contribution of the high flux population ($S>S_0\simeq 1$\,mJy) to the measured cosmic radio background, $\int_{S_0} dS \, S \, (dN/dS) / B_{CRB}$, is about $\simeq 16$ per cent at $1.4$\,GHz, and the corresponding surface number density $N(>S_0)=\int_{S_0} dS \, (dN/dS) \simeq 4.7 \times 10^{5}$\,sr$^{-1}$.  One could ask if a significant portion of the flux from the high flux population has been missed by radio surveys. If this were the case, then the contribution of the high flux population to the measured background would have been underestimated. One possible source of the missing flux could be  extended low surface brightness sources. However, as mentioned above, the surveys included in Figure\,\ref{scbfigure}, in spite of the fact that they span a wide range of resolutions and are obtained from different interferometer arrays, show very good agreement on the integrated contribution to the background.  We will return to the contribution of low surface brightness sources and quantify the possible missed flux further in \S\,\ref{mlsb}.

\subsection{Low-Flux Population} \label{lf}

Unlike the high-flux population, the total contribution of the low-flux population is not fixed by existing surveys, since these do not constrain the low flux peak on the $S^2 (dN/dS) - S$ plot. Even though at low fluxes ($S < S_0$) source counts indicate a power-law distribution $dN / dS \propto S^{-\gamma}$, the measured faint end index $\gamma$ varies substantially between different surveys, ranging from $\gamma =2.11$ claimed by \citet{Fomalont02} and $\gamma=2.61$ claimed by \citet{OM08}. The value of $\gamma=2.5$ is what would be seen in a  static, isotropic and homogeneous  Euclidean universe. In extrapolating to lower fluxes, as long as the faint end index is above 2, the integrated contribution from lower flux sources to the background will continue to increase.  Obviously, at some eventual lower flux, the faint end index must drop below 2, to avoid Olbers' paradox.

\begin{figure}
\includegraphics[width=3.5in]{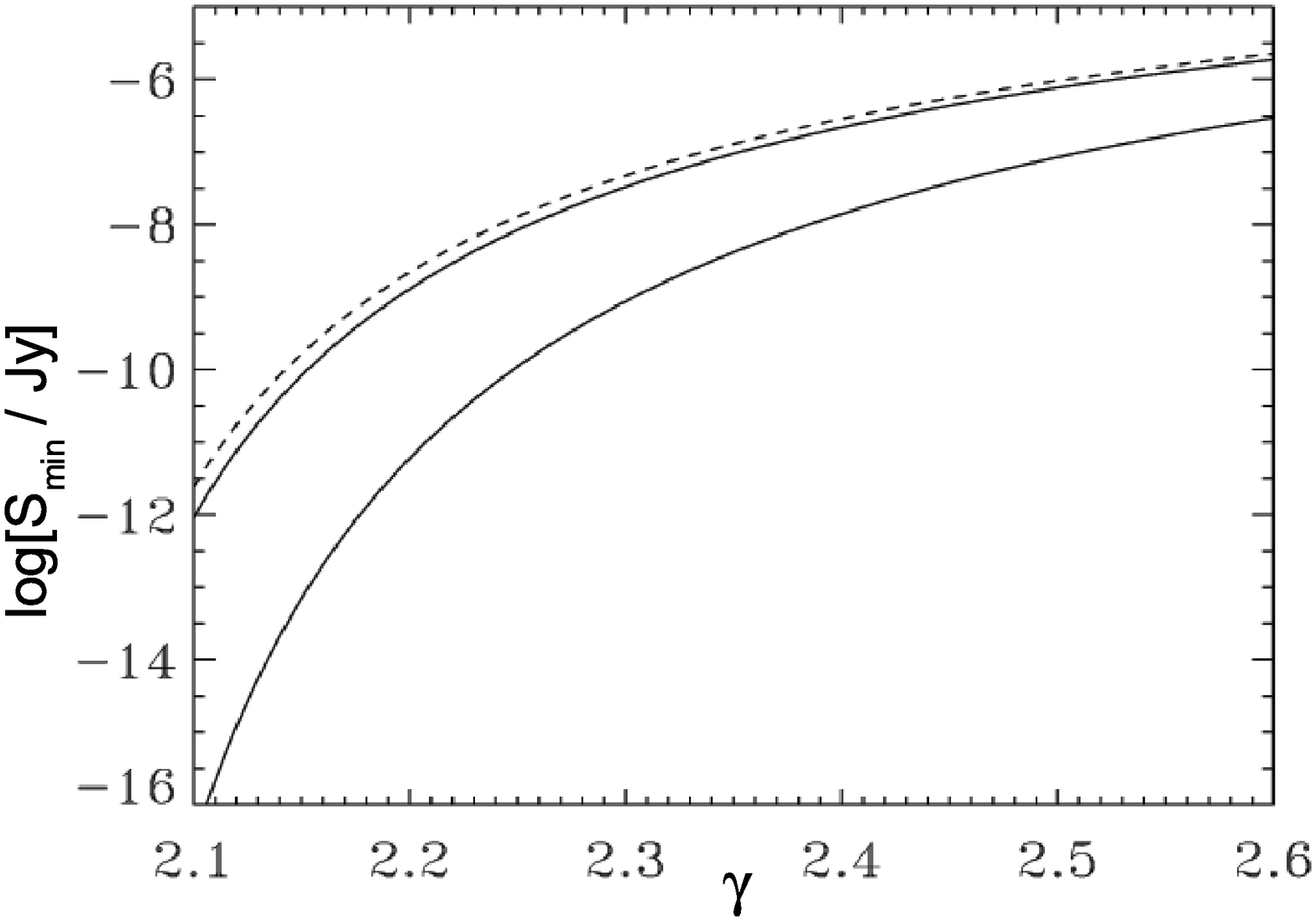}
\includegraphics[width=3.5in]{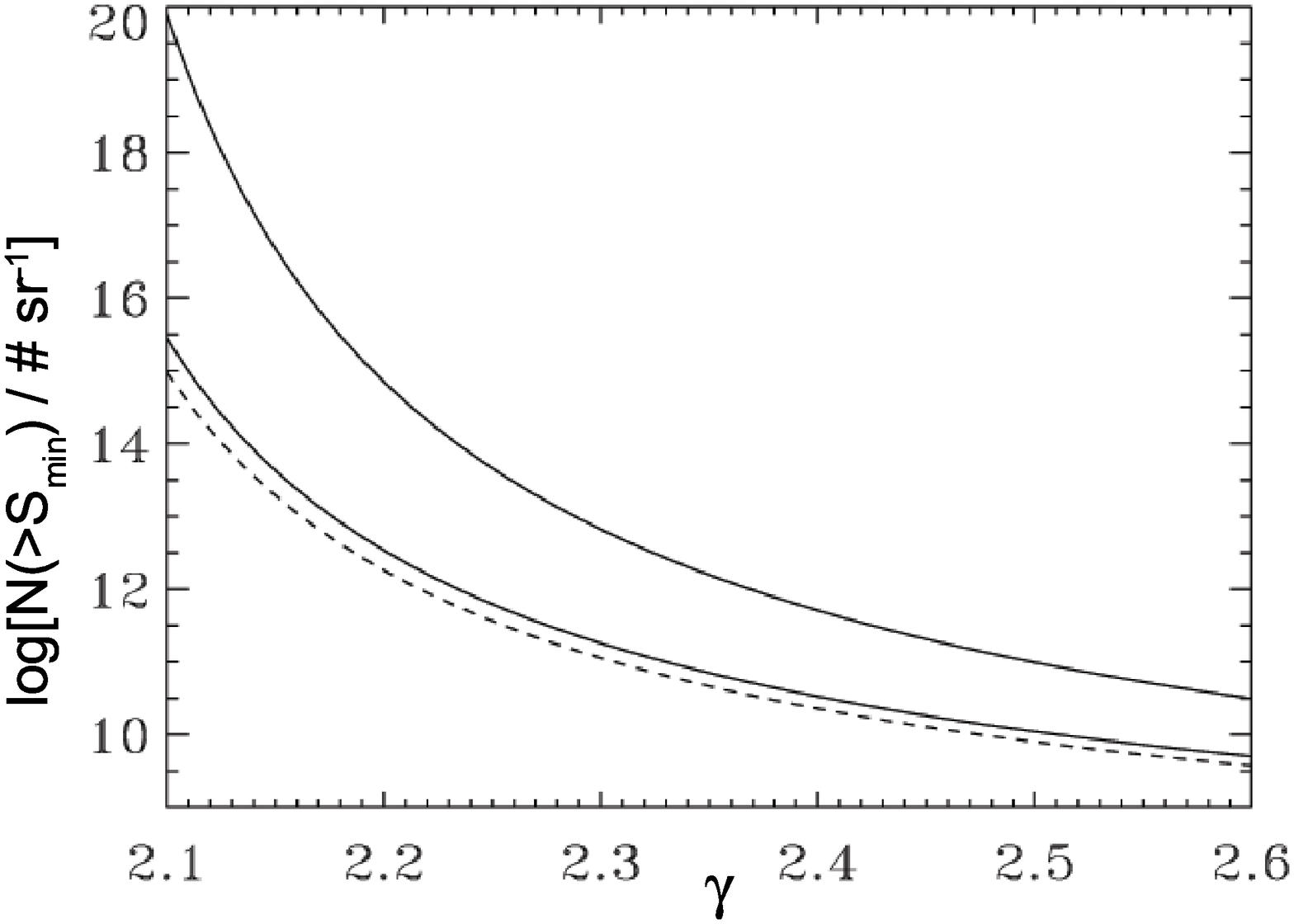}
\caption{{\bf TOP:} The lower flux density of sources $S_{min}$ at 1.4 GHz required so that the extrapolated source counts from $S_{0}\simeq 1$\,mJy, with a faint end index $\gamma$, can account for observed CRB as reported by the ARCADE 2 collaboration, including a 16 per cent (solid lines) or 25 per cent (dashed line; see \S\,\ref{mlsb}) contribution to the background from S$\geq$S$_{0}$ sources.  The upper and lower solid curves are for the two extreme values of the 1 mJy normalizations n$_{0}$ of the surveys we consider. For $\gamma \sim$ 2.5 an extrapolation to $S_{min} \sim 0.1 \mu$Jy is needed, while the needed lower flux density falls rapidly with decreasing faint end index. {\bf BOTTOM:} Same as the top panel, but for the minimum number of sources below S$_{0}$ needed to produce the observed CRB.}
\label{fluxlimitfigure}
\end{figure}

Assuming 
\begin{equation}
{dN \over dS} = k_0 \left({S \over S_0}\right)^{-\gamma} \, \quad {\rm for} \quad \, S_{\rm min}<S<S_0, 
\end{equation}
one can calculate the required value of $S_{\rm min}$ as a function of $\gamma$ such that the sources in this range provide the rest of the CBR. It is easy to show that for $\gamma>2$
\begin{equation}
S_{min} = S_0 \, \left[ {(1-H) \, (\gamma - 2) \, B_{CRB} \over k_0 S_0^2 } + 1\right]^{1/(2-\gamma)} \, , 
\end{equation}
\noindent
where the factor $H$ is the estimated fractional contribution from sources above $S_0\simeq 1$\,mJy, which we determine to be $0.16$.  The top panel of Figure\,\ref{fluxlimitfigure} shows the variation of $S_{\rm min}$ with $\gamma$ at $\nu= 1.4$ GHz. We can also estimate the corresponding minimum number of required sources to be
\begin{equation}
N\left(>S_{min}\right)-N(>S_0) = {k_0 S_0^{1-\gamma} \over \gamma - 1} \left[\left({S_{min} \over S_0}\right)^{1-\gamma} - 1\right] \, , 
\end{equation}
whose dependence on $\gamma$ is shown on the bottom panel of Figure\,\ref{fluxlimitfigure}.  Thus, if the background is to be made primarily from low flux sources, their faint end index below $1$\,mJy should be close to $\gamma = 2.5$ in order to not exceed reasonable estimates for the total number of non-dwarf galaxies in the obervable Universe.  In other words, in order for low flux density sources to account for the observed CRB, they must reach low flux values of less than 1 $\mu$Jy and have a surface density of ($\ga 10^{10}$\,sr$^{-1}$), but not much higher than this value.  

A modeling of the dN/dS distribution of the low-flux ($<$ mJy) population in terms of a single power law, although customary, is more than likely an over-simplification.  A more realistic description would involve a smoothly curved function with the integrated contribution per log flux bin (Figure \ref{scbfigure}, lower panel) peaking around $S_{min}$ and then falling off, but including some contribution from objects with fluxes lower than $S_{min}$.  The large population of dwarf galaxies may be the sources that increasingly become relevant at fluxes below $S_{min}$.

\section{Diffuse Sources} \label{xrb}

In this section we consider the possibility that the CRB results from truly diffuse emission associated with the large scale structure of the Universe, such as the intracluster medium (ICM), intergalactic medium (IGM), or the filaments connecting the clusters containing warm-hot gas.  First we consider very general constraints, and then look at specific possibilities.

\subsection{General constraints on diffuse emission}

The simplest constraint on any population of electrons producing the CRB is that it must have a relatively flat energy spectrum, in order to produce the observed spectral index of $\alpha \sim 0.6$. Less obviously, we show below that it must be associated with a magnetic field of at least 1 $\mu$G. This is because otherwise the magnetic field energy density in such systems is much less than the energy density of the cosmic microwave background (CMB) and other background radiations. As a result, the relativistic electrons responsible for the radio emission would lose most of their energy producing hard X-ray and gamma radiation via inverse Compton scattering of the other background fields. We now derive this limit more carefully.

Given the observed power law spectrum, the energy density of the radio background per frequency dex at $\nu_r$ is given by
\begin{equation}
\label{obs}
[\nu_r U_{\nu_r}] = {4 \pi \over c} \, [\nu_r B_{CRB}\!(\nu_r)] = 1.17 \, {8 \pi k_B \nu_{\star}^3 \over c^3} \, \left({\nu_r \over \nu_{\star}}\right)^{0.4} \, , 
\end{equation}
\noindent
where $\nu_{\star} = 1$\,GHz.  We are considering the background spanning frequencies from $\nu_1\,\sim$1 MHz to $\nu_2\,\sim$10 GHz.

For a density of ultrarelativistic electrons with a power-law energy spectrum
\begin{equation}
 n_e(\gamma_e) = k_e \, \gamma_e^{-s} \quad {\rm for} \quad \gamma_{e_{1}} \leq \gamma_e \leq \gamma_{e_{2}}\, , 
\end{equation}
\noindent
where $k_{e}$ is a normalization constant in units of cm$^{-3}$, and $\gamma_{e_{1}}$ and $\gamma_{e_{2}}$ correspond here to the Lorentz factors of the electrons producing radiation primarily around $\nu_1$ and $\nu_2$, respectively, the synchrotron emissivity may be approximated as \citep[see e.g.][]{ryb79}
\begin{equation}\label{model}
[\nu_r j_{\nu_r}] \simeq {c \sigma_T U_B \over 6 \pi} \, k_e \left({ \nu_r \over \nu_{cr}}\right)^{{3-s \over 2}} \, , 
\end{equation}
\noindent
where $U_B \equiv B^2 / 8 \pi$ is the energy density of the magnetic field, and $\nu_{cr}\, \equiv (3eB/4\pi mc) \, \gamma_{e}^2 \, \simeq 4.2 \, (B/ \mu{\rm G}) \, \gamma_{e}^2 \,$ Hz is the critical (radio) synchrotron frequency for a given $\gamma_{e}$.   

For production of the observed radio background we need $s=2.2$, $\gamma_{e_{1}}=5\times 10^2 (B/ \mu{\rm G})^{-1/2}$, and $\gamma_{e_{2}}=5\times 10^4 (B/ \mu{\rm G})^{-1/2}$, with the value of $k_e$ being determined from the following relation between the emissivity and the observed energy density. 

We relate the (radio) synchrotron energy density to the emissivity with 
\begin{eqnarray}\label{emis}
[\nu_r U_{\nu_r}] & = & {4\pi \over c} \int {dV \over dz} {dz \over 4\pi d_L^2(z)} [\tilde{\nu}_r \, j_{\tilde{\nu}_{r}}] = \nonumber \\ & = & [\nu_r \, j_{\nu_{r}}]\, {4\pi \over H_0} \,  \int {F_{syn}(z) \, dz \over (1+z)^{(s+1)/2} E(z)} \, , 
\end{eqnarray}
where $\tilde{\nu}_r = \nu_r\,(1+z)$, $H_0=70$ km$^{-1}$ Mpc$^{-1}$ is the Hubble constant, and $F_{syn}(z)$ describes the evolution of the product $U_B\times k_e$.  Here $E(z) \equiv \sqrt{\Omega_M \, (1+z)^3 + \Omega_{\Lambda}}$ for the assumed flat cosmology, and the comoving volume element is
\begin{equation}\label{volume}
 {dV \over dz} \equiv {c^3 4 \pi \over H_0^3 E(z)} \, \left[\int_0^z {dz' \over E(z')}\right]^2 = {4\pi c \, d_L^2 \over H_0 \, (1+z)^{2} \, E\!(z)} \, . 
\end{equation}

Similarly, we can get the inverse Compton (IC) emissivity resulting from the same population of electrons as\footnote{The following expressions are valid in the Thompson regime and are a good approximation for scattering of photons with $h\nu_0 \sim  1$ ev by the highest energy electrons $\gamma_2$. For scattering of photons above this energy one must use the Klein Nishina cross section.  Few relevant photons lie above this range so in what follows we approximate the Klein-Nishina suppression by a sharp cutoff.}
\begin{equation}
[\nu_{ic} j_{\nu_{ic}}] = {3 c h \sigma_{\rm T} \over 16 \pi} \, \int d \nu_0 \, \int d \gamma_e \, n(\nu_0) \, n_{e}(\gamma_e) \, {\nu_{ic}^2 f_T \over \gamma_e^2 \, \nu_0} \, , 
\end{equation}
\noindent
\citep{Blum70}, where $n(\nu_0) \equiv [\nu_0 U_{\nu_0}]_{tot} / h \nu_0^2$ is the total spectral number density of the extragalactic background photon field (including radio, CMB, infrared, optical/UV, X-ray, etc),
\begin{eqnarray}
& & f_{T} = 2 q_T \, \ln q_T + q_T + 1 - 2 q_T^2 \, , \nonumber \\  & & q_T \equiv {\nu \over 4 \nu_0 \gamma_e^2} \quad {\rm and} \quad {1 \over 4 \gamma_e^2} \leq q_T \leq 1 \, . 
\end{eqnarray}
\noindent

The IC emissivity can be related to the IC energy density by an equation similar to equation \ref{emis} with $F_{syn}(z)$ replaced by $F_{ic}(z)$ which now describes the evolution of the product $n(\nu_0)\times k_e$.  Ignoring the differences between these two evolutions (i.e. assuming that the ratio of the integrals over redshift involving $F_{syn}(z)$ and $F_{ic}(z)$ is of order unity) and by eliminating $k_e$ we can express the IC energy density in terms of the synchrotron energy density as
\begin{eqnarray}\label{fin}
[\nu_{ic} U_{\nu_{ic}}] \simeq {[\nu_r U_{\nu_r}] \over B^2/8 \pi} \, \left({4.2 \, (B/\mu{\rm G}) \over \nu_{\star}}\right)^{0.4} \, \nu^2 \, \times \nonumber  \\ \times \int_{\gamma_{e_{1}}(B)}^{\gamma_{e_{2}}(B)} d\gamma_e \, \int_{\nu_{min}}^{\nu_{max}} d\nu_0 \, \nu_0^{-3} \gamma_e^{-4.2} \, [\nu_0 U_{\nu_0}]_{tot}  \, f_T \, ,
\end{eqnarray}
\noindent
with $\nu_{max} = \min[\nu_2, \,\, \nu, \,\, m_ec^2 / 4 h \gamma_e]$ and $\nu_{min} = \max[\nu_1, \,\, \nu/4 \gamma_e^2]$.  We note that because the IC emissivity is dominated by upscattering of the CMB photon field, for which the energy density increases with redshift, the presented evaluation of $[\nu_{ic} U_{\nu_{ic}}]$ with the cosmological evolution neglected corresponds strictly to a lower limit. 

\begin{figure}
\includegraphics[width=3.5in]{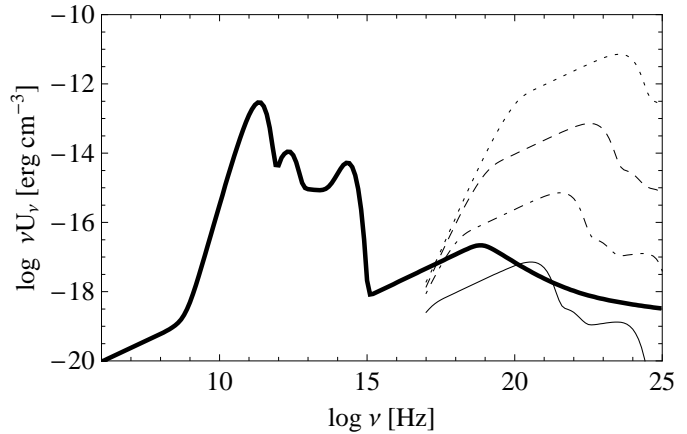}
\caption{The thick black curve shows the measured energy density of the radio, microwave, infrared, optical, ultraviolet, X-ray, and $\gamma$-ray extragalactic backgrounds. The other curves show the energy density produced by inverse Compton scattering of the photon backgrounds by electrons necessary to produce the radio background reported by the ARCADE 2 collaboration via synchrotron emission. The dotted, dashed, dot-dashed and solid curves are for a $1$\,nG, $10$\,nG, $l00$\,nG, and $1$\,$\mu$G level average magnetic field. Because the intergalactic magnetic field is known to be $\leq$ 1$\mu$G, the observed level of the X-ray background rules out a significant portion of the radio background being produced by electrons far from galaxies. Spectral energy densities of the IR/optical, X-ray, and $\gamma$-ray cosmic photon fields were constructed to be in agreement with the background levels provided by \citet{fra08}, \citet{gil07}, and \citet{sre98}, respectively.}
\label{bgndfigure}
\end{figure}

Figure\,\ref{bgndfigure} shows the energy density of the observed extragalactic background light (thick curve), and the expected IC energy density resulting from upscattering of these background photons by electrons producing the radio background as given by equation \ref{fin}, for different magnetic fields $B = 0.001$, $0.01$, $0.1$, and $1$\,$\mu$G (dotted, dashed, dot-dashed, and solid curves, respectively).  Spectral energy densities of the IR/optical, X-ray, and $\gamma$-ray cosmic photon fields were constructed to be in agreement with the background levels provided by \citet{fra08}, \citet{gil07}, and \citet{sre98}, respectively.  Clearly any magnetic field weaker than 1$\mu$G would result in X-ray/$\gamma$-ray emission exceeding the observed background, and regions with such magnetic fields may be excluded as significant sources of the CRB.

Importantly, this consideration excludes our own Galactic halo as the origin of the bulk of the isotropic radio signal.  \citet{Taylor09} use rotation measure measurements of 37,000 polarized extragalactic radio sources to determine the intensity of the magnetic field in the Galactic halo, concluding with a value of approximately 1$\mu$G.  In our Galactic halo, the level of the ambient optical and infrared photon fields will be even higher than that considered in the calculation here, by an amount depending on the distance from the Galactic plane, predicting an X-ray background many times larger than that observed. There are Galactic and solar system components to the observed diffuse X-ray background, but these are significant only below 1 keV \citep{HM06}. The observed level of the X-ray background therefore strongly disfavours a Galactic origin for the observed isotropic radio signal.

\subsection{Diffuse emission from the IGM}

The CRB could in principle result from a population of relativistic electrons pervading the IGM as as a whole, perhaps resulting from many generations of AGN. Large scale radio sources can easily expand to sizes of many Mpc on gigayear timescales, and so overlap. Almost certainly, however, adiabatic expansion losses would result in a very low energy density, and synchrotron losses would lead to a steep energy spectrum. Furthermore a variety of arguments suggest that the magnetic field in the IGM is likely to be very weak, $B \leq 0.2$\,$\mu$G (see \citet{val04} and references therein). Diffuse emission from the IGM therefore seems unlikely to be the solution.

\subsection{Diffuse emission from clusters}

There is evidence for diffuse radio emission in some but not all clusters of galaxies \citep[e.g.][]{fer08}. Ever since discovery of this emission from the Coma cluster \citep{Large59}, there have been improved observations of this system \citep{Willson70,Schlick87}, as well as more extensive surveys by \citet{GF00} who have identified more than 40 clusters with diffuse radio emission. These are classified as halos or relics, and are often associated with dynamically disturbed (merging) clusters. The origin of the relativistic electrons is controversial. They may be injected by AGN jets \citep[and references therein]{mcn07} or produced as secondary pairs from ultrarelativistic cosmic-ray hadrons scattering of ICM protons. Finally, non-thermal electrons within IGM may be also picked up directly from the thermal pool and accelerated to ultrarelativistic energies by ICM shocks or turbulence \citep[`primary' electrons; see e.g.,][]{pet01,BL06}. We note that the most recent observations of few clusters at $\gamma$-ray energies seem to exclude the `secondary' nature of the radio halo electrons \citep[e.g.,][]{ale10}.  

The fraction of clusters with radio emission increases rapidly with increasing soft X-ray luminosity, indicating that {\em strong} diffuse radio emission is not a common property of all clusters. However, deeper radio observations, especially at low frequencies, suggest that the presence of extended, low surface-brightness radio structures in galaxy clusters is relatively common \citep[e.g.][]{rud09} but at lower flux levels. Possibly then the integrated effect of a large number of such weakly emitting clusters could be a significant contribution to the CRB. We are not able to predict the level of this integrated emission, but we can check the cluster hypothesis against our two general constraints -- magnetic field and spectral index.

There are estimates of the magnetic fields in some clusters via Faraday rotation measurements, which indicate (line-of-sight averaged) magnetic fields of 1 to 10 $\mu$G \citep[e.g.,][]{Kim90,Taylor02}. In particular, in Coma, \citet{Kronberg07} measure a line-of-sight averaged intergalactic magnetic field in the range $0.2-0.4$\,$\mu$G. 
However, the actual magnetic field will be larger by a factor of $(R/H_B)^{1/2}$ if it is tangled or chaotic on a distance scale $H_B$) that is smaller than cluster size $R$. For $H_B\sim 10$ kpc one can get fields as high as few $\mu$G. Overall, it is quite plausible that some cluster emission passes the magnetic field test.

However, the observed radio spectra are relatively steep (index $\alpha >1$; see \citet{Liang00}) which does not match the index of the CRB. It therefore seems that diffuse ICM/IGM emission, while possibly widespread, is not a dominant contributor to the cosmic radio background.

\section{Missing Low-Surface Brightness Emission} \label{mlsb}

It is possible that the calculated strength of the CRB based on source counts is underestimated in two ways, due to the surface brightness limits of surveys.  Firstly, the fluxes of some extended objects may have been undercharacterized, if their sizes are large, or they have extended low surface brightness components. Secondly, if there is a wide range of source surface brightnesses extending to low values, there may exist sources which will not be detected.  In order to discuss this issue in more detail, it is useful to consider high and low flux sources separately.

\subsection{High-Flux Sources}

At high fluxes, source counts are dominated by radio galaxies and radio-loud quasars that very often have large scale morphology. It is useful to distinguish the two well known Fanaroff-Riley morphological types \citep{fan74}. The low-power FR\,Is have surface brightness profiles which fade outwards with no clear end, and therefore part of their total flux will be undetected in interferometric radio surveys.  However, these outer components typically have steep spectra \citep[e.g.,][] {Hardcastle99,lai06,lai08}, which is inconsistent with being the dominant component of the CRB.  Moreover, they have only a modest cosmological evolution \citep{wil01,jam04,rig08,smo09} and so although significant at sub-mJy fluxes \citep[see][] {rig08,pad09}, are unlikely to be the dominant low-flux population.\footnote{This conclusion is in agreement with the modest estimated contribution of FR\,I sources to the extragalactic X-ray and $\gamma$-ray backgrounds \citep[][respectively]{cel04,sta06}}

`Classical doubles' (FR\,IIs), on the other hand, have clearly defined boundaries of their extended radio lobes, but are expected to decrease in surface brightness as they age and grow in size \citep{Kaiser97}.  Such extended structures, which can reach Mpc-scale sizes, may to fall below typical survey surface brightness detection limits, especially at high redshifts.  Even the cores may remain undetected, as they often have very low levels of central (AGN-like) activity \citep{mac01,mac06,dwa09}.  However, the known low-surface brightness giant radio galaxies tend to have steep spectra \citep[see][]{jam08,kon08}, which is not consistent with that of the CRB.  

A new survey, the Australia Telescope Low Brightness Survey \citep{Sub08,Sub09} has addressed the question of the number of potentially missed low-surface brightness radio galaxies or their extended components, using an $8.4$\,deg$^{2}$ survey down to mJy fluxes, with a surface brightness threshold a factor of 5 lower than previous comparable surveys. \citet{Sub08} state that 30 per cent of their sources have at least half of their flux between $5$\,arcsec and $30$\,arcsec, and that 10 per cent of their sources have a size at least 1.5 times their beamsize of $50$\,arcsec, indicating that there is a significant, but not dominant, component of extended flux.  This survey discovered a few new giant radio galaxies.  The authors have not yet published source counts, but their initial analysis finds 500 sources in $8.4$\,deg$^{2}$ to a completeness limit of $1$\,mJy at $1.4$\,GHz, corresponding to an integrated source density of $1.95 \times 10^5$\,sr$^{-1}$. This is not very different from our estimate based on other surveys discussed in \S\,\ref{hf}, suggesting that any missing population of high flux but low surface brightness sources is not large enough to explain the CRB.

In summary, while there are indications that some high flux density, low surface brightness sources may have been missed in earlier source count surveys, it seems unlikely that this is a very large effect; at most a 50 per cent correction, which can increase the maximum integrated contribution of the high flux population to the CRB to $\sim$ 25 per cent.

\subsection{Low-Flux Sources}

Below $1$\,mJy, optical spectra of identified sources indicate that source counts become dominated by star forming galaxies \citep[e.g.,][]{Benn93} and the characteristic radio source size changes from $\sim 10$\,arcsec to $\sim 2$\,arcsec \citep{CC85}, consistent with a change in population from double-lobed, elliptical-hosted radio galaxies to starforming `regular' spiral galaxies and Seyferts. Sizes of a few arcsecond are large enough that measured peak fluxes will often be an underestimate of total flux, but not by a large factor. The usual procedure is to apply a statistical correction to the fluxes, assuming a given source size distribution according to the formula of \citet{Windhorst90}.

For example, \citet{Hopkins03}, with a beam size of 6 arcsec, apply the \citet{Windhorst90} formula for the angular flux distibution to estimate a correction factor of $f \sim 1.3$ for an assumed median source size of $2$ arcsec.  A median size of $4$\,arcsec (considerably bigger than expected for spiral galaxies at the redshift $z=1$) would instead require $f \sim 1.7$. This effect is therefore unlikely to be very large for the star forming galaxy population.

One could imagine a population of large and low surface brightness radio-emitting galaxies whose fluxes are badly underestimated. Then for median source sizes of $10$ and $50$\,arcsec, the correction factors would be $f \sim 2.5$ and $f \sim 5.9$, respectively, the later being nearly sufficient to explain the observed CRB. However, a completely missing population at $S \sim\,0.1$\,mJy seems unlikely. In compiling the source counts shown in Figure\,\ref{scbfigure}, we examined normalization against survey resolution, and found no significant consistent effect. There is also some clear evidence in a specific field. With a beam size of $6$\,arcsec, \citet{CC85} surveyed a VLA field to $\mu$Jy levels which had previously been surveyed at the same frequency with a beam size of $19$\,arcsec by \citet{CM84}. Of 159 sources found in the first (large beam) survey, only 8 were not seen in the later (small beam) survey.

Summarizing the situation for low flux sources, it seems unlikely that significant extended flux has been missed (i.e. not already corrected for).

\section{Point Sources} \label{nc}

The above results indicate that a class of radio sources not previously considered is needed to produce the bulk of the radio background. In particular, the background cannot be formed from diffuse emission or relatively few high luminosity sources, but rather many lower luminosity sources. As discussed in \S\,\ref{lf}, these sources should be numerous to enough dominate source counts from below the $\sim 10$\,$\mu$Jy limit of current radio surveys down to the $\sim 10^{-2}$\,$\mu$Jy level. In addition, the sources should have high magnetic fields, $B > 1$\,$\mu$G, and should be produced by a relatively flat-spectrum of relativistic electrons. In this section we consider some possible candidates.

\subsection{Radio Supernovae}\label{rsn}

Radio supernovae (RSNe)produce significant radio flux and could be a contributor to the background. RSNe associated with Type\,II supernovae have a mean spectral index roughly comparable to the observed CRB \citep{Weiler86}.  \citet{Colina01} discovered a RSN in a radio monitoring campaign of the galaxy NGC\,7469 with a $8.4$\,GHz luminosity of $\simeq 1.1 \times 10^{21}$\,W\,Hz$^{-1}$, which is more than a thousand times more luminous than Cassiopeia\,A, the brightest radio supernova remnant in the Milky Way, and which would correspond to flux densities in the range $\mu$Jy\,$< S < $\,mJy between $z=0.1$ and $z=2$.  This object would not have been found in conventional optical searches. 

Are RSNe of this luminosity numerous enough to account for the background?  As discussed in \citet{Weiler04} significant radio emission is only seen in core collapse supernovae, and approximately 10 per cent of core collapse supernovae exhibit radio loudness. \citet{Madau98} estimate the number of core collapse supernovae integrating across all redshifts to be $3$\,arcmin$^{-2}$\,yr$^{-1}$. \citet{Mannucci07} update this with the number of supernovae obscured optically by starbursts, limiting the correction to a factor of two at most, giving $\leq 1 \times 10^{9}$\,yr$^{-1}$ core collapse supernovae over the whole sky.  With the period of radio loudness lasting on the order of a year \citep{Weiler04}, we have only $\leq 1 \times 10^{8}$\, RSNe on the sky at any time.  Noting that $z=2$ is the peak era of supernova activity and assuming a mean RSNe 8.4\,GHz luminosity of $1\times 10^{21}$\,W\,Hz$^{-1}$, this would give a contribution to the background intensity at the level of $\la 10$\,Jy/sr, falling short of the observed CRB by a factor of $\geq 1000$.  For a broad luminosity function extending below $1\times 10^{21}$\,W\,Hz$^{-1}$, the contribution will be even less.  Note also that a fraction of the RSNe population at this luminosity located at significantly lower redshifts would have sufficiently high flux to be included in the radio surveys considered in \S\,\ref{sourcecounts} and Figure \ref{scbfigure}, and therefore cannot explain the 'missing' portion of background.  

A higher fraction of core collapse supernovae could be more luminous in the radio if they were exploding in denser environments.  At very high redshifts, nearly all population III stars could be very massive and explode in so called pair-instability supernovae.  However, the supernova rate from population III stars is estimated by \citet{WA05} to be only $10^{-4}$\,arcmin$^{-2}$\,yr$^{-1}$, which is much less than regular core collapse supernovae.  

Finally, if RSNe were significant contributors to the radio fluxes of spiral galaxies, they would be seen as compact sources in high-resolution images of nearby spiral galaxies. However, they are not. Young RSNe can sometimes be seen as individual sources, but they contribute only a tiny fraction of the total flux.  We conclude that RSNe are not a significant component of the radio background.

\subsection{Radio Quiet Quasars}\label{rq}

Radio quiet (RQ) quasars, as is well known, are not radio silent, and thus could potentially account for a missing population of CRB sources.  They are typically unresolved on sub arcsecond scales, with $1.4$\,GHz luminosities of the order of $10^{38}-10^{41}$\,erg\,s$^{-1}$ \citep{BK07,White07,Elvis94}. The radio emission of higher luminosity RQ quasars is most often produced by mildly/non-relativistic electrons in nuclear outflows originating in the inner parts of accretion disks, though large-scale relativistic jets are sometimes (surprisingly) detected in such systems \citep{blu03}.  The expected strong magnetization of disk winds ($B \gg 1$\,$\mu$G) requires low Lorentz factors and low energies of radio-emitting electrons which remove the IC problem discussed in \S\,\ref{xrb}.  We note that the radio emission from most low-luminosity ($< 10^{22}$ W/Hz) RQ quasars may not originate in nuclear outflows, but rather from star-forming host galaxies \citep{Kellerman94}, but their space density is much lower than that of comparably luminous sources in normal starforming galaxies.  The observed radio continua of RQ quasars are flat, possibly in agreement with the CRB spectrum (Blundell, priv.com.).  RQ quasars are also numerous, though underrepresented in existing radio surveys due to the flux limits of the surveys.  In addition, \citet{White07} have demonstrated that stacking seemingly empty radio images of optically identified quasars which are individually below the noise level of the radio survey leads to a composite image of significant radio flux.  Finally, \citet{Simpson06}, \citet{Ibar09} and \citet{pad09} have claimed recently that radio quiet AGN constitute a significant fraction of the sub-mJy radio source population.

On the other hand, the idea that RQ quasars produce the bulk of the CRB presents several problems.  From Figure\,\ref{fluxlimitfigure}, even with a faint end index of $2.5$ below $100$\,$\mu$Jy, the number of sources needed to make the background is perhaps only a factor of 10 lower than reasonable estimates of the total number of non-dwarf galaxies in the observable Universe, and as such is much higher than the expected number of quasars, i.e. high-accretion rate objects. Also, optical quasar number counts \citep[e.g.][]{Richards06} point to a peak in the contribution to $S^{2} dN/dS$ occurring at $\sim 50$\,$\mu$Jy in $i$ band.  Converting this to a radio flux using a radio loudness parameter (the ratio of the 1.4 GHz radio to 2500\,\AA\, optical luminosity) on the order of 1 for RQ quasars, and adjusting the optical flux between $i$ band and 2500\,\AA\, according to standard spectral models, indicates the peak of $S^{2} dN/d\log(S)$ at $1.4$\,GHz would fall near $10$\,$\mu$Jy. Below this peak flux the faint end index would drop to a sub-Euclidean value. This turn over would occur at least one order of magnitude higher than that necessitated by Figure\,\ref{fluxlimitfigure}.

Let us now estimate the expected contribution of RQ quasars to the measured CRB. Knowing the \emph{radio} luminosity function (LF) of RQ quasars, $\psi (L_R, z)$, where $L_R \equiv [\nu_r L_{\nu_r}]$ is the radio luminosity at frequency $\nu_r$ (in what follows we will use $\nu_r=5$ GHz), their contribution to the CRB energy density can be evaluated in a manner equivalent to that of equation (\ref{emis}) with the emissivity replaced by the LF multiplied by $L$ or $L_R\psi(L_R,z)$, namely
\begin{eqnarray}\label{rqed}
[\nu_r U_{\nu_r}] & = & {1 \over c} \, \int dz \int dL_R \, {dV \over dz} \, {\psi(L_R, z) \,L_R \over (1+z)^{\alpha-1} 4 \pi \, d_L^2} = \nonumber \\ & = & \int dz \int dL_R \,\,{\psi (L_R, z) \, L_R \over H_0\,(1+z)^{1+\alpha}\,E(z)} \, , 
\end{eqnarray}
\noindent
where $\alpha = d\ln L_{\nu_r}/d\ln \nu_r$ is the radio spectral index.

We do not have direct knowledge of the radio LF but we can estimate as described below by relating it to the known optical \emph{bolometric} LF of quasars $\psi (L_{\rm bol}, z)$ which may evolve as a pure luminosity evolution, pure density evolution, or luminosity-dependent density evolution as constrained by \citet{hop07}. Assuming similar evolution for the radio LF, then by definition,
\begin{equation}
\psi (L_R, z) \, dL_R =\psi (L_{\rm bol}, z) \, dL_{\rm bol} \, .
\end{equation}
\noindent
To evaluate the above integral we need the relation between the bolometric and radio luminosities. \citet{White07} give a relation between the radio luminosity at 5\,GHz and the absolute $2500$\,\AA\, magnitude $M_{UV}$, which can be written as $\log \left[L_{\rm 5\,GHz}/(\rm erg\,s^{-1}\,Hz^{-1})\right] \simeq 22.07 - 0.34 M_{UV}$ or $\log \left[L_R/(\rm erg\,s^{-1})\right]= 31.76 - 0.34 \, M_{UV}$. We convert the 2500\,\AA\, magnitude to the bolometric quasar luminosity using the approximate relations $L_{\rm bol} \simeq 3 \times \nu_{UV}L_{\nu_{UV}}$, $m_{UV} - M_{UV} = 5 \, \log [d_L/{\rm pc}] - 5+K$, where $K$ stands for the $K$-correction factor, and $m_{UV} = -2.5 \, \log [S_{UV}/3500\,{\rm Jy}]$, to obtain
\begin{equation}
\left( {L_R \over 10^{36}\,{\rm erg\,s^{-1}}} \right) \simeq 0.9\times \left( {L_{\rm bol} \over 10^{41}\,{\rm erg\,s^{-1}}}\right)^{0.85} \, ,
\end{equation}
\noindent
where the luminosities $L_{\rm bol}$, $\nu_{UV}L{\nu_{UV}}$, and $L_R$ are expressed in the units of erg\,s$^{-1}$.

After integrating equation \ref{rqed} over the redshift range\footnote{While some high accretion rate objects may be present before z=6, the number must necessarliy fall far short of the number of objects needed to make the background outlined in \S\,\ref{lf}, as the number of galaxies present at these redshifts is a small fraction of the total in the observable universe.} $z = 0-6$ and the radio luminosity range $L_R = 10^{36}-10^{42}$\,erg\,s$^{-1}$ (corresponding to the bolometric luminosity range $L \simeq 10^{41}- 10^{48}$\,erg\,s$^{-1}$), for the fit parameters regarding the quasar luminosity function as provided in \citet{hop07}, we find the expected contribution of RQ quasars to the extragalactic radio background at $5$\,GHz to be at the level of $\simeq$ 1.4 to 1.7 per cent for pure luminosity or luminosity-dependent density evolution models. In the case of a pure density evolution model, the expected contribution rises to 4 per cent. This contribution is therefore smaller by at least a factor of $10$ than needed. One has to keep in mind, however, that the presented calculations correspond strictly to the lower limits, because in constructing the radio luminosity function, we considered only type 1 quasars. However, including the population of type 2 sources, which cannot outnumber the population of unobscured quasars by a large factor, is not expected to account for the order-of-magnitude difference.

In order for emission from RQ quasars to comprise the bulk of the background, therefore, their radio loudness must evolve with redshift, and they must be significantly more numerous at redshifts between 1 and 3 than previously thought. Although \citet{Blundell03} has claimed that significant flux from RQ quasars may be missed by interferometric observations, we conclude that emission from RQ quasars, while significant (at the few percent level at least), does not likely explain the bulk of the missing contribution to the radio background, unless there is some as of yet undetected but significant evolution in the quasar radio luminosity function.

\subsection{Starforming Galaxies} \label{gals}

Given the large ($\ga 10^{10}$\,sr$^{-1}$) number of sources needed to produce the CRB, the constraints on extended and non-galactic emission discussed in sections \ref{mlsb} and \ref{xrb}, the required strong magnetic fields ($B > 0.2$\,$\mu$G)  and flat spectra, and  the shortcomings of radio supernovae and RQ quasars (\S\,\ref{rsn} and \S\,\ref{rq}), it seems that ordinary galaxies may be more reasonable candidates for production of most of the CRB. Ordinary star forming galaxies are thought to dominate the low-flux population of the resolved radio sources \citep[e.g.][]{Richards00}. The radio continua produced in star forming regions are relatively flat, being characterized by the spectral indices close to the one of the CRB $\alpha = 0.6$ \citep[see][]{Ibar09}.  Finally, the expected magnetic fields $B \ga 1$\,$\mu$G in star forming regions  circumvents the IC problem discussed in \S\,\ref{xrb}.

\subsubsection{Constraints from the far-infrared background} \label{sf}

The radio emission originating in local star forming regions is observed to be correlated with the  far-infrared (FIR) flux \citep[e.g.][and references therein]{DB02,Ibar08}, which combined with the  observed level of the infrared background, constrains the contribution of these sources to the CRB.  \citet{DB02} give a relation between the 1.4 GHz radio power $P_{\rm 1.4\,GHz}$ and the FIR luminosity, $L_{FIR}$, in the wavelength range $10-1000$\,$\mu$m, which can be written as $L_r\equiv [\nu  P(\nu)]_{1.4 {\rm GHz}}\sim \times 10^{-6}\, L_{FIR}$.  Therefore, if the radio background is produced  in the star forming regions of Seyferts, spirals, and ULIRGs, then from a naive application of this  ratio one should expect the observed FIR background at the level of
\begin{displaymath}
[\nu U_{\nu}]_{FIR} \simeq 10^6 \, [\nu U_{\nu}]_{\rm 1.4\,GHz} \simeq 2 \times 10^{-13} \, {\rm  erg\,cm^{-3}} \, ,
\end{displaymath}
\noindent
which is higher than that observed \citep{Marsden09} by a factor of $>10$. This is in agreement with  \citet{DB02} result showing that the expected surface brightness of the sky at $178$\,MHz for  different models of the cosmologically evolving star formation rate to be between $3$\,K and $30$\,K;  about 3 to 30 per cent of  the ARCADE result  $T_{CRB}(\rm 178\,MHz) \simeq 104$\,K \citep{Fixsen09}.

An alternative way to formulate this problem is as follows.  \citet{Appleton04} find that individual  starforming galaxies locally show a ratio between FIR and radio fluxes given by the value $q_{70}   \equiv \log(S_{\rm 70\,\mu m} / S_{\rm 1.4 GHz}) = 2.15$.  If we take the observed CRB and assume it is made by star forming galaxies following this ratio, then, ignoring for now any K-correction issues, we predict a FIR surface brightness at  $70$\,$ \mu$m of $178$\,nW\,m$^{-2}$\,sr$^{-1}$, which is about $25$ times the observed level reported  in \citet{Dole06}.  Thus we can conclude that the contribution of systems that obey the local radio FIR correlation to the radio background is on the level of approximately $\la 5$ per cent, or, alternately,  that the radio background must be made in large part from sources that appear more radio loud than the local radio FIR correlation by a factor of approximately 20.

\subsubsection{An evolving radio far-infrared correlation} \label{evolve}

Since star formation in the Universe has evolved and the rate was much higher at redshifts of 1 and above \citep[e.g.,][]{Madau98}, the bulk of both the radio and infrared backgrounds are produced at these redshifts.  Therefore, an evolution in the observed FIR to radio flux ratio towards greater radio loudness with redshift is necessary to have the radio and FIR backgrounds produced at the observed levels by starforming galaxies. 

It is important to note that the ratio of the relative contribution to the FIR background and radio background surface brightnesses from a galaxy (or class of galaxies) located at a particular redshift is given by the non-K-corrected FIR to radio flux ratio, rather than the K-corrected ratio that is often reported for higher redshifts.  Relating an observed flux ratio or ratio evolution to a  K-corrected one, or vice-versa, is dependent on the source SED assumed, and the wavelengths in question.  For observations at $70$\,$\mu$m and $1.4$\,GHz and a typical starburst galaxy SED, a fixed intrinsic $q_{70}$ (where $q_x$ is the log of the flux ratio at infrared frequency x to 1.4 GHz radio) from a galaxy at $z\sim 2$ will show an observed value reduced by $\Delta q\sim 0.6$ (see example calculations in references below). To explain the observed discrepancy between radio and $70$\,$\mu$m backgrounds shown in \S \ref{sf}, we would need to observe a high redshift evolution of $\Delta q({\rm observed})\sim -1.3$ or $\Delta q({\rm intrinsic})\sim -0.7$.

A number of papers have explored the evolution of $q$ with redshift at either MIR, FIR or submm wavelengths.  These determinations are complicated by selection effects, and in the case of reports of the K-corrected correlation, template spectra.  Although some papers conclude that there is no evidence for a change in intrinsic $q$ (eg \citet{Ibar08}, \citet{Sargent10}, \citet{Ibar08}) others do find possible evolution in intrinsic $q$ (eg \citet{Vlahakis07}, \citet{Beswick08}, \citet{Seymour09}, \citet{Bourne10}), although it is not clear that this is large enough to explain the radio background.  Recent data from the BLAST instrument \citep{Ivison10} also show the intrinsic correlation evolving, and new results from the Herschel instrument \citep{2Ivison10} are consistent with this as well.  The most suitable reported results for comparison are those of \citet{Sargent10} and \citet{Bourne10} who use Spitzer $70$\,$\mu$m data and present the evolution in observed $q_{70}$ versus redshift (Fig 12 in Sargent et al, Fig 9 in Bourne et al).  These figures agree in showing that by $z\sim 2$ a change of $\Delta q({\rm observed})\sim -0.7$ is observed, i.e. a factor of 5 decline in the FIR to radio flux ratio.  While significant, this is not enough to explain the FIR/radio background discrepancy.  However, these figures show considerable scatter, and Sargent et al show that a radio-selected, as opposed to IR-selected, sample shows a systematically lower value, with $\Delta q({\rm observed})\sim -0.4$.  We note again that determinations of the value of the correlation at higher redshifts are complicated by selection biases and contamination by AGN emission.  

An evolving radio to FIR luminosity ratio toward the radio loud with increasing redshift would suggest that one or more of the following is true at higher redshifts: 1) a larger portion of the energy of star formation goes into in relativistic particles, for some reason possibly related to a different structure of the interstellar medium shaping the cosmic-ray acceleration efficiency at shocks driven by supernova,  2) a larger portion of the stars are high mass, resulting in more supernovae as well as an enhanced cosmic-ray acceleration at shocks driven by winds from massive stars,  3) synchrotron emissivity is enhanced by higher interstellar magnetic fields, or 4) AGN activity in ordinary galaxies is proportionally more important.

There is no concrete information on item 1, while item 2 requires evolution of the initial mass function, which if true would alter the interpretation of the data on the star formation rate.  The third possibility can be neither rejected nor supported by current data.

Here let us only comment on the possibility that the cosmological evolution of supermassive black  holes in the galactic centers results in enhanced AGN activity in late-type galaxies at high  redshifts, and thus in their enhanced radio emission relative to infrared.  Note that in this context that it is now established that a large fraction of nearby spiral galaxies show weak AGN activity  \citep[]{Ho97,Ho08}. In addition, several authors have argued that supermassive black holes (SMBHs) were spinning  more rapidly at early epochs \citep{Wang09}. For spiral-hosted SMBHs the effect may be even stronger than for the elliptical hosted SMBHs that give rise to quasars, due to the different character of the dominant accretion events which determine the black hole spin evolution \citep[see the discussion in][]{sik07,Volonteri07}. In the framework of the spin paradigm for jet production, critically re-examined recently by \citet{sik07}, one could therefore expect more powerful jets, and therefore more AGN-related radio emission (for the same accretion luminosity) at higher redshifts in these systems. 

If the radio background is indeed formed from $\sim 0.1$\,$\mu$Jy sources, that corresponds to about $\sim 10^{21}$\,W\,Hz$^{-1}$ radio ($1.4$\,GHz) luminosity at $z=2$, which would be between $10^{8}$  and $10^{9}$ $L/L_{\sun}$ at $70$\,$\mu$m with an evolving (K-corected) $q_{70}$.  This infrared luminosity is below the 'knee' of the infrared luminosity function, i.e. the radio background would be made by relatively normal spiral galaxies, in contrast to the infrared background, which appears to be dominated by luminous and ultraluminous IR galaxies \citep{Magnelli09}.  We note the difficulty that locally the peak contribution to radio emission is from relatively luminous galaxies with 1.4 GHz radio luminosities of around $10^{22}$\,W\,Hz$^{-1}$ \citep{Condon02}, while the peak contribution to the radio background, from galaxies beyond redshift 1, is at 1.4 GHz luminosities of around $10^{21}$\,W\,Hz$^{-1}$.

\section{Discussion}\label{discussion}

We have considered several mechanisms as the origin of the 'missing' radio flux necessary to account for the radio background, assuming it is at the level reported by the ARCADE 2 collaboration.  As shown in \S\,\ref{xrb}, diffuse, low surface brightness synchrotron emisssion from large scale structures (IGM, ICM and WHIM), and from our own Galactic halo, is limited by the observed level of the X-ray/$\gamma$-ray background.  As discussed in \S\,\ref{hf}, radio sources detected in many surveys above $1$\,mJy at 1.4 GHz contribute a total of 16 per cent of the background, and considerations of possible missed flux from extended sources discussed in \S\,\ref{mlsb} may increase this number by at most a factor of 1.5. The resolved sources with fluxes from below $1$\,mJy to the current lower limit of interferometric surveys at $\sim 10$\,$\mu$Jy contribute another $\sim 10$ per cent, leaving between 60 per cent and 75 per cent of the cosmic radio background unaccounted for.

We find it difficult to explain the level of the radio background reported by ARCADE 2 without a new population of low flux radio sources.  These sources should be numerous and faint enough to dominate source counts below the $\sim 10$\,$\mu$Jy limit of current radio surveys and must extend to the $\sim 10^{-2}$\,$\mu$Jy (at 1.4 GHz) level.  Moreover, they should have an observed ratio of radio to infrared output a factor of 5 above what is observed in local galaxies. As discussed in \S\,\ref{rsn} and \S\,\ref{rq}, radio supernovae and radio quiet quasars do not seem to produce adequate radio emission to account for the bulk of the background, although the latter class of objects is expected to contribute to the CRB at the level of at least a few percent.  We conclude that the for the radio background to be at the level reported by ARCADE 2, it must be largely comprised of emission from ordinary galaxies at $z > 1$ in which the radio to far-infrared observed flux ratio increases significantly with redshift.

\section*{Acknowledgments}

We thank Philip Best, Katherine Blundell, Chi C. Chueng, Sebastian Jester, Dan Marrone, Matthew Turk, and Tom Abel for their input.  JS thanks the ARCADE team, S. Kahn, and R. Schindler for their encouragement.  \L S was supported by the Polish Ministry of Science and Higher Education through the project N N203 380336, and also by the scandinavian NORDITA program on `Physics of Relativistic Flows'.

\clearpage

\begin{table*}%[tbph]
\begin{minipage}{140mm}
%\begin{deluxetable}{rcrccrrr}
%\scriptsize
%\begin{center}
\caption{Radio source count data plotted in Figure \ref{scbfigure}}
\label{sctable}
%\tablecolumns{8}
\begin{tabular}{rrrrrrrr}
%\tablewidth{5.5 in}
%\startdata
Survey & Frequency & Facility$^{a}$  & Resolution$^{b}$ & S$_{limit}$$^{c}$ & $\gamma$$^{d}$ & k$^{d}$  & $S_{max}$$^{e}$ \\
  & (GHz) &  & (arcsec) & ($\mu$Jy) &  &  & (mJy)  \\ \hline
\citet{Fomalont02}$^{f}$ & 8.4 & VLA & 6 & 7.5 & 2.11 $\pm$ .13$^{g}$ & 16.8 & 1\\
 &  &  & 3.5 & 35.0 &  & \\
\citet{Windhorst93} & 8.4 & VLA & 10 & 14.5 & 2.3 $\pm$ .2$^{g}$ & 4.6 & 1 \\
\citet{Ciliegi03} & 5 & VLA & 4 & 50 & 2.0$^{g}$  & 131.7 & 20 \\
\citet{Prandoni06} & 5 & ATCA & 11 & 400 & 1.93$^{h}$  & 289.7 & 33.5 \\
\citet{Gruppioni97} & 2.7 & ATCA & 39 & 400 & 1.58$^{h}$  & 2173.8 & 21.5 \\
\citet{Richards00} & 1.4 & VLA & 2 & 40 & 2.38 $\pm$ .13$^{g}$   & 8.25  & 1 \\
\citet{Hopkins03} & 1.4 & ATCA & 6$\times$12 & 60 & 1.83 ($\leq$6 mJy)$^{i}$  & - & 1000 \\
 &  &  &  &  & 1.89 ($\geq$6 mJy)  & - & \\
\citet{Bondi03} & 1.4 & VLA & 31 & 60 & 2.28 ($\leq6$ mJy)$^{g}$  & 27.3 & 40 \\
 &  &  &  &  & 1.79 ($\geq$6 mJy) & 1062.3 &  \\
\citet{Windhorst84} & 1.4 & WSRT & 12.5 & 600 & 1.9 ($\leq$1 mJy)$^{i}$ & 341 & 10000 \\
\citet{Windhorst85} & 1.4 & VLA & 14.7 & 225 & 2.1 ($\leq1$ mJy)$^{h}$ & 175.3 & 100 \\
 &  &  &  &  & 1.9 ($\geq$1 mJy)  &  867.5 & \\
\citet{Fomalont06} & 1.4 & VLA A & 1.8 & 25 & 2.43$^{g}$  & 9.2 & 10000 \\
\citet{OM08} & 1.4 & VLA A & 1.6 & 15 & 2.61$^{h}$ & 2.0 & 1500 \\
\citet{Simpson06} & 1.4 & VLA C & 5$\times$4 & 100 & 1.9$^{h}$ & 505.2 & 10000 \\
\citet{BI06}$^{f}$ & 1.4 & VLA A & 1.6 & 15 & 2.24$^{h}$ & 33.1 & 1153 \\
 & & & 1.5 & 48 & 2.37$^{h}$ & 12.9 & 1327 \\
 & & & 1.5 & 29 & 2.44$^{h}$ & 7.2 & 854 \\
\citet{Ibar09} & 1.4 & VLA B & 4.3$\times$4.2 & 20 & 2.08$^{h}$ & 38.4 & 3000 \\
\citet{Ibar09} & .610 & GMRT & 7.1$\times$5.6 & 45 & 2.09$^{h}$ & 284.4 & 1000 \\
\citet{Bondi07} & .610 & VLA & 6 & 150 & 1.84$^{h}$ & 1346.2 & 200 \\
\citet{KM85} & .610 & WSRT & 57$\times$57 & 9000 & 1.94$^{h}$ & 1158 & 900 \\
\citet{WV82} & .610 & WSRT & 60 & 10000 & 1.94$^{g}$  & 1039 & 516 \\
\citet{VR89} & .408 & PSRT & 204$\times$276 & 70000 & 1.9 $\pm$ .1$^{g}$   & 2200 & 516 \\
\citet{Grueff88} & .408 & NCTRT & 156$\times$288 & 70000 & 2.28$^{h}$ & 1072 & 10000 \\
\citet{PK78} & .408 & CMBR & 80 & 10000 & 1.8$^{h}$ & 2284 & 10000 \\
\citet{Hales88} & .151 & CMBR & 300 & 120000 & 1.9 ($\leq$1 Jy)$^{h}$ & 3824.3 & 20000 \\
 &  &  & &  & 2.3 ($\geq$1 Jy) & 3421.2 &  \\
\citet{MCG90} & .151 & CMBR & 70$\times$100 & 80000 & 2.07 ($\leq$1 Jy)$^{h}$ & 3777.4 & 10000 \\
 &  &  & &  & 2.68 ($\geq$1 Jy) & 3874.5 &  \\
\hline
\end{tabular}
%\enddata
$^{a}$VLA= Very Large Array, ATCA= Australia Telescope Compact Array, WSRT=Westerbrook Synthesis Radio Telescope, PSRT=Penticon Synthesis Radio Telescope, NCTRT=Northern Cross Transit Radio Telescope, CMBR= Cambridge Radio Telescope Survey.\\
$^{b}$The resolution quoted here is the full width at half max of the composite image.  If the beam is not round, values for two axes are given.\\
$^{c}$The low flux limit of the survey is a factor determined by the survey authors (usually 5) times the RMS noise.\\
$^{d}$From fits to $dN/dS$ of the form $dN/dS = k\times \,S^{-\gamma}$, where the $dN/dS$ distribution is expressed in sr$^{-1}$ Jy$^{-1}$, and the flux $S$ in Jy.\\
$^{e}$The highest flux object observed, or the highest flux listed in a $dN/dS$ table or fit to $dN/dS$.\\
$^{f}$Separate results for different fields reported.\\
$^{g}$Source counts power law directly given.\\
$^{h}$$dN/dS$ given in table and power law determined with fit.\\
$^{i}$Source counts given by equation in $S$.\\
%\end{center}
\end{minipage}
\end{table*}
%\end{deluxetable}

\label{lastpage}


\begin{thebibliography}{}
\bibitem[\protect\citeauthoryear{Aleksi{\'c} et al.}{2010}]{ale10} Aleksi{\'c} J., et al., 2010, ApJ, 710, 634
\bibitem[\protect\citeauthoryear{Appleton et al.}{2004}]{Appleton04} Appleton P. et al., 2004, ApJS, 154, 147
\bibitem[\protect\citeauthoryear{Ballantyne}{2009}]{bal09} Ballantyne D.R., 2009, preprint (arXiv:0904.0996)
\bibitem[\protect\citeauthoryear{Benn et al.}{1993}]{Benn93} Benn C., Rowan-Robinson M., McMahon R., Broadhurst T., Lawrence A., 1993, MNRAS, 263, 98
\bibitem[\protect\citeauthoryear{Beswick et al.}{2008}]{Beswick08} Beswick R., Muxlow T., Thrall, H. Richards A., Garrington S., 2008, MNRAS, 385, 1143
\bibitem[\protect\citeauthoryear{Biggs \& Ivison}{2006}]{BI06} Biggs A., Ivison R., 2006, MNRAS, 371, 963
\bibitem[\protect\citeauthoryear{Blumenthal \& Gould}{1970}]{Blum70} Blumenthal G.R., Gould R.~J., 1970, Reviews of Modern Physics, 42, 237
\bibitem[\protect\citeauthoryear{Blundell}{2003}]{Blundell03} Blundell K.~M., 2003, New Astronomy Review, 47, 593
\bibitem[\protect\citeauthoryear{Blundell \& Kuncic}{2007}]{BK07} Blundell K., Kuncic Z., 2007, ApJ, 668, 103
\bibitem[\protect\citeauthoryear{Blundell et al.}{2003}]{blu03} Blundell K.M., Beasley A.~J., Bicknell G.~V., 2003, ApJL, 591, L103
\bibitem[\protect\citeauthoryear{Bondi et al.}{2003}]{Bondi03} Bondi M. et al., 2003, A\&A, 403, 857
\bibitem[\protect\citeauthoryear{Bondi et al.}{2007}]{Bondi07} Bondi M. et al., 2007, A\&A, 463, 519
\bibitem[\protect\citeauthoryear{Bourne et al.}{2010}]{Bourne10} Bourne L., Dunne L., Ivison R., Maddox S., Dickinson M., Frayer D., 2010, MNRAS, submitted, preprint (arXiv:1005.3115)
\bibitem[\protect\citeauthoryear{Brunetti \& Lazarian}{2006}]{BL06} Brunetti G., Lazarian A., 2006, IAUJD, 1, 13
\bibitem[\protect\citeauthoryear{Celotti \& Fabian}{2004}]{cel04} Celotti A., Fabian A.C., 2004, MNRAS, 353, 523
\bibitem[\protect\citeauthoryear{Ciliegi et al.}{2003}]{Ciliegi03} Ciliegi P., Zamorani G., Hasinger G., Legmann I, Szokoly G., Wilson G., 2003, A\&A, 398, 801
\bibitem[\protect\citeauthoryear{Coleman \& Condon}{1985}]{CC85} Coleman P., Condon J., 1985, AJ, 90, 1431
\bibitem[\protect\citeauthoryear{Colina et al.}{2001}]{Colina01} Colina L., Alberdi A., Torrelles J., Panagia N., Wilson S., 2001, ApJL, 553, L19
\bibitem[\protect\citeauthoryear{Condon, Cotton, \& Broderick}{2002}]{Condon02} Condon J., Cotton W., Broderick J., 2002, AJ, 124, 675
\bibitem[\protect\citeauthoryear{Condon}{2007}]{Condon07} Condon J., 2007, ASP conference series, volume 80, eds: Afonso, J. Ferguson, C. Mobasher, \& B. Norris, R. p. 189
\bibitem[\protect\citeauthoryear{Condon \& Mitchell}{1984}]{CM84} Condon J., Mitchell K., 1984, AJ, 87, 1429
\bibitem[\protect\citeauthoryear{Dole et al.}{2006}]{Dole06} Dole H., et al. 2006, A\&A, 451, 417
\bibitem[\protect\citeauthoryear{Dwarakanath \& Kale}{2009}]{dwa09} Dwarakanath K.S., Kale R., 2009, ApJL, 698, 163
\bibitem[\protect\citeauthoryear{Dwek \& Barker}{2002}]{DB02} Dwek E., Barker M., 2002, ApJ, 575, 7D
\bibitem[\protect\citeauthoryear{Elmouttie et al.}{1995}]{elm95} Elmouttie E., Haynes R.~F., Jones K.~L., Ehle M., Beck R., Wielebinski R., 1995, MNRAS, 275, L53  
\bibitem[\protect\citeauthoryear{Elvis et al.}{1994}]{Elvis94} Elvis M. et al., 1994, APJS, 95, 1
\bibitem[\protect\citeauthoryear{Fanaroff \& Riley}{1974}]{fan74} Fanaroff B.L., Riley J.M., 1974, MNRAS, 167, 31P
\bibitem[\protect\citeauthoryear{Feretti \& Giovannini}{2008}]{fer08} Feretti L., Giovannini G.\, 2008, in Pilonis, M., Lopez-Cruz, O., Hughes, D., eds, Lecture Notes in Physics vol. 470, A Pan-Chromatic View of Clusters of Galaxies and the Large-Scale Structure, Springer-Verlag, Berlin, p. 143
\bibitem[\protect\citeauthoryear{Fixsen et al.}{2010}]{Fixsen09} Fixsen D. et al., 2010, ApJ, submitted, preprint (arXiv:0901.0559)
\bibitem[\protect\citeauthoryear{Fomalont et al.}{2002}]{Fomalont02} Fomalont E., Kellerman K., Partridge R., Windhorst R., Richards E., 2002, ApJ, 123, 2402
\bibitem[\protect\citeauthoryear{Fomalont et al.}{2006}]{Fomalont06} Fomalont E., et al., 2006, ApJS, 167, 103
\bibitem[\protect\citeauthoryear{Franceschini et al.}{2008}]{fra08} Franceschini A., Rodighiero G., Vaccari M., 2008, a\&A, 487, 837
\bibitem[\protect\citeauthoryear{Gervasi et al.}{2008}]{Gervasi08} Gervasi M., Tartari A., Zannoni M., Boella G., Sironi G., 2008, ApJ, 682, 223
\bibitem[\protect\citeauthoryear{Gilli et al.}{2007}]{gil07} Gilli R., Comastri A., Hasinger G., 2007, A\&A, 463, 79
\bibitem[\protect\citeauthoryear{Giovannini \& Feretti}{2000}]{GF00} Giovannini G., Feretti L., 2000, New Astronomy Review, 5, 335
\bibitem[\protect\citeauthoryear{Gruppioni et al.}{1997}]{Gruppioni97} Gruppioni C., Zamorani G, de Ruiter H., Parma P., Mignoli M, Lari C., 1997, MNRAS, 286, 470
\bibitem[\protect\citeauthoryear{Grueff}{1988}]{Grueff88} Grueff G., 1988, A\&A, 193, 40
\bibitem[\protect\citeauthoryear{Hales et al.}{1988}]{Hales88} Hales S., Baldwin J., Warner P., 1988, MNRAS, 234, 919
\bibitem[\protect\citeauthoryear{Hardcastle}{1999}]{Hardcastle99} Hardcastle M., 1999, A\&A, 349, 381
\bibitem[\protect\citeauthoryear{Hickox \& Markevitch}{2006}]{HM06} Hickox R., Markevitch M., 2006, ApJ, 645, 95
\bibitem[\protect\citeauthoryear{Ho}{2008}]{Ho08} Ho L., 2008, ARAA, 46, 475
\bibitem[\protect\citeauthoryear{Ho, Filippenko, \& Sargent}{1997}]{Ho97} Ho L., Filippenko A., Sargent W., 1997, Proceedings of the IAU colloquium No. 159, eds. Peterson, B. Cheng, F.- Z. and Wilson, A. p. 429
\bibitem[\protect\citeauthoryear{Hopkins et al.}{2003}]{Hopkins03} Hopkins A., Alfonso J., Chan B., Cram L., Georgakakis A., Mobasher R., 2003, AJ, 125, 465
\bibitem[\protect\citeauthoryear{Hopkins et al.}{2007}]{hop07} Hopkins P.F., Richards G.T., Hernquist L., 2007, ApJ, 654, 731
\bibitem[\protect\citeauthoryear{Hummel, van Gorkom, \& Kotanyi}{1983}]{hum83} Hummel E., van Gorkom J.~H., Kotanyi C.~G., 1983, ApJ, 267, L5 
\bibitem[\protect\citeauthoryear{Ibar et al.}{2008}]{Ibar08} Ibar E., et al., 2008, MNRAS, 386, 953
\bibitem[\protect\citeauthoryear{Ibar et al.}{2009}]{Ibar09} Ibar E., Ivison R., Biggs A., Lal D., Best P., Green D., 2009, MNRAS, 397, 281
\bibitem[\protect\citeauthoryear{Ivison et al.}{2010a}]{Ivison10} Ivison R. et al., 2010a, MNRAS, 402, 245
\bibitem[\protect\citeauthoryear{Ivison et al.}{2010b}]{2Ivison10} Ivison B., et al. 2010b, A\&A, 518, 31
\bibitem[\protect\citeauthoryear{Jamrozy}{2004}]{jam04} Jamrozy M., 2004, A\&A, 419, 63
\bibitem[\protect\citeauthoryear{Jamrozy et al.}{2008}]{jam08} Jamrozy M., Konar C., Machalski J., Saikia D.J., 2008, MNRAS, 385, 1286
\bibitem[\protect\citeauthoryear{Kaiser et al.}{1997}]{Kaiser97} Kaiser C., Dennett-Thorpe J., Alexander P. 1997, MNRAS, 292, 723
\bibitem[\protect\citeauthoryear{Katgert-Merkelijn et al.}{1985}]{KM85} Katgert-Merkelijn J., Robertson J., Windhorst R., Katgert P., 1985, A\&AS, 61, 517
\bibitem[\protect\citeauthoryear{Kellerman et al.}{1994}]{Kellerman94} Kellerman K., Sramek R., Schmidt M., Green R., Shaffer D. 1994, AJ, 108, 1163
\bibitem[\protect\citeauthoryear{Kim \& Kronberg}{1990}]{Kim90} Kim K., Kromberg P., 1990, ApJ, 355, 29
\bibitem[\protect\citeauthoryear{Kogut et al.}{2010}]{Kogut09} Kogut A. et al., 2010, ApJ, submitted, preprint (arXiv:0901.0562)
\bibitem[\protect\citeauthoryear{Konar et al.}{2008}]{kon08} Konar C., Jamrozy M., Saikia D.J., Machalski, J., 2008, MNRAS, 383, 525
\bibitem[\protect\citeauthoryear{Kronberg et al.}{2007}]{Kronberg07} Kronberg P., Kothes R., Salter C., Perillat P., 2007, ApJ, 659, 257
\bibitem[\protect\citeauthoryear{Liang et al.}{2000}]{Liang00} Liang H., Hunstead R., Birkinshaw M., Andreani P., 2000, ApJ, 544, 686
\bibitem[\protect\citeauthoryear{Liang et al.}{2006}]{lai06} Liang R.A., Canvin J.R., Cotton W.D., Bridle, A.H., 2006, MNRAS, 368, 48
\bibitem[\protect\citeauthoryear{Liang et al.}{2008}]{lai08} Liang R.A., Bridle A.H., Parma P., Feretti L., Giovannini G., Murgia M., Perley R.A., 2008, MNRAS, 386, 657
\bibitem[\protect\citeauthoryear{Large et al.}{1959}]{Large59} Large M., Mathewson D., Haslam C., 1959, Nature, 183, 1663L
\bibitem[\protect\citeauthoryear{Machalski et al.}{2001}]{mac01} Machalski J., Jamrozy M., Zola S., 2001, A\&A, 371, 445
\bibitem[\protect\citeauthoryear{Machalski et al.}{2006}]{mac06} Machalski J., Jamrozy M., Zola S., Koziel D., 2006, A\&A, 454, 85
\bibitem[\protect\citeauthoryear{Madau et al.}{1998}]{Madau98} Madau P., Della Valle M., Panagia N., 1998, MNRAS, 297, L17
\bibitem[\protect\citeauthoryear{Magnelli et al.}{2009}]{Magnelli09} Magnelli B., Elbaz D., Chary R., Dickinson M., LeBorgne D., Frayer D. Willmer C., 2009, A\&A, 496, 57
\bibitem[\protect\citeauthoryear{Mannucci et al.}{2007}]{Mannucci07} Mannucci F., Della Valle M., Panagia N., 2007, MNRAS, 377, 1229
\bibitem[\protect\citeauthoryear{Marsden et al.}{2010}]{Marsden09} Marsden G. et al., 2010, ApJ, 707, 1729
\bibitem[\protect\citeauthoryear{McGilchrist et al.}{1990}]{MCG90} McGilchrist M., Baldwin J., Riley J., Titterington D., Waldram E., Warner P., 1990, MNRAS, 246, 110
\bibitem[\protect\citeauthoryear{McNamara \& Nulsen}{2007}]{mcn07} McNamara B.R., Nulsen, P.E.J., 2007, ARAA, 45, 117
\bibitem[\protect\citeauthoryear{Owen \& Morrison}{2008}]{OM08} Owen F., Morrison G., 2008, ApJ, 136, 1889
\bibitem[\protect\citeauthoryear{Padovani et al.}{2009}]{pad09} Padovani P., Mainieri V., Tozzi P., Kellermann K.I., Fomalont E.B., Miller N., Rosati P., Shaver P., 2009, ApJ, 694, 235
\bibitem[\protect\citeauthoryear{Pearson \& Kus}{1978}]{PK78} Pearson T. Kus A., 1978, MNRAS, 182, 273
\bibitem[\protect\citeauthoryear{Petrosian}{2001}]{pet01} Petrosian V., 2001, ApJ, 557, 560
\bibitem[\protect\citeauthoryear{Prandoni et al.}{2006}]{Prandoni06} Prandoni L., Parma P., Wieringa M., de Ruiter H., Gregorini L., Mignano A., Vettolani G., Ekers R., 2006, A\&A, 457, 517
\bibitem[\protect\citeauthoryear{Richards}{2000}]{Richards00} Richards E., 2000,  ApJ, 533, 611
\bibitem[\protect\citeauthoryear{Richards et al.}{2006}]{Richards06} Richards E. et al.,, 2006,  AJ, 131, 2766
\bibitem[\protect\citeauthoryear{Rigby et al.}{2008}]{rig08} Rigby E.E., Best P.N., Snellen I.A.G., 2008, MNRAS, 385, 310
\bibitem[\protect\citeauthoryear{Rudnick \& Lemmerman}{2009}]{rud09} Rudnick L., Lemmerman J.A., 2009, ApJ, 697, 1341
\bibitem[\protect\citeauthoryear{Rybicki \& Lightman}{1979}]{ryb79} Rybicki G.B., Lightman A.P., 1979, {\it `Radiative Processes in Astrophysics'}, New York: Wiley 1979
\bibitem[\protect\citeauthoryear{Sargent et al.}{2010}]{Sargent10} Sargent M., et al., 2010, ApJ, 186, 341 submitted, preprint (arXiv:1005.1072)
\bibitem[\protect\citeauthoryear{Schlickeiser et al.}{1987}]{Schlick87} Schlickeiser R., Sievers A., Thiemann H., 1987, A\&A, 182, 21
\bibitem[\protect\citeauthoryear{Seiffert et al.}{2010}]{Seiffert09} Seiffert M. et al., 2010, submitted, preprint (arXiv:0901.0559)
\bibitem[\protect\citeauthoryear{Seymour et al.}{2009}]{Seymour09} Seymour N. et al. 2009, MNRAS, 2009, 398, 1573
\bibitem[\protect\citeauthoryear{Sikora et al.}{2007}]{sik07} Sikora M., Stawarz {\L}., \& Lasota J.P., 2007, ApJ, 658, 815
\bibitem[\protect\citeauthoryear{Singal et al.}{2010}]{Singal09}  Singal J. et al. 2010, ApJ, accepted, preprint (arXiv:0901.0546)
\bibitem[\protect\citeauthoryear{Simpson et al.}{2006}]{Simpson06}  Simpson C. et al. 2006, MNRAS, 372, 741
\bibitem[\protect\citeauthoryear{Smol{\v c}i{\'c} et al.}{2009}]{smo09} Smol{\v c}i{\'c} V. et al., 2009, ApJ, 696, 24
\bibitem[\protect\citeauthoryear{Sreekumar et al.}{1998}]{sre98} Sreekumar P., et al., 1998, ApJ, 494, 523
\bibitem[\protect\citeauthoryear{Stawarz et al.}{2006}]{sta06} Stawarz {\L}., Kneiske T.~M., Kataoka J., 2006, ApJ, 637, 693
\bibitem[\protect\citeauthoryear{Subrahmanyan et al.}{2010}]{Sub08}  Subrahmanyan R., Ekers R., Saripalli L., Sadler M., 2010, MNRAS, 402, 2792
\bibitem[\protect\citeauthoryear{Subrahmanyan et al.}{2009}]{Sub09}  Subrahmanyan R. et al., in prep
\bibitem[\protect\citeauthoryear{Taylor et al.}{2002}]{Taylor02}  Taylor G., Fabian A., Allen S., 2002, MNRAS, 334, 769
\bibitem[\protect\citeauthoryear{Taylor, Stil, \& Sunstrum.}{2009}]{Taylor09}  Taylor A., Stil J., Sunstrum C., 2009, ApJ, 702, 1230
\bibitem[\protect\citeauthoryear{Vall{\'e}e}{2004}]{val04} Vall{\'e}e J.P., 2004, New Astronomy Review, 48, 763
\bibitem[\protect\citeauthoryear{Vall{\'e}e \& Roger}{1989}]{VR89} Vallee J., Roger R., 1989, A\&AS, 77, 31
\bibitem[\protect\citeauthoryear{Vlahakis et al.}{2007}]{Vlahakis07} Vlahakis C., Eales S., Dunne L., 2007, MNRAS, 379, 1042
\bibitem[\protect\citeauthoryear{Volonteri et al.}{2007}]{Volonteri07} Volonteri M., Sikora M., Lasota J., 2007, ApJ, 667, 704
\bibitem[\protect\citeauthoryear{Wang et al.}{2009}]{Wang09} Wang J. et al., 2009, ApJL, 697, 704
\bibitem[\protect\citeauthoryear{Weiler et al.}{1986}]{Weiler86} Weiler K., Sramek R., Panagia N., van der Hulst J., Salvati M., 1986, ApJ, 301,  790
\bibitem[\protect\citeauthoryear{Weiler et al.}{2004}]{Weiler04} Weiler K., van Dyk S., Sramek R., Panagia N., 2004, new astro. reviews, 09.017
\bibitem[\protect\citeauthoryear{White et al.}{2007}]{White07} White R., Helfand D., Becker R., Glickman E., De Vries W., 2007, ApJ,  654,  99
\bibitem[\protect\citeauthoryear{Willott et al.}{2001}]{wil01} Willott C.J., Rawlings S., Blundell K.M., Lacy M., Eales S.A., 2001, MNRAS, 322, 536
\bibitem[\protect\citeauthoryear{Wilson}{1970}]{Willson70} Wilson M., 1970, MNRAS, 151, 1
\bibitem[\protect\citeauthoryear{Wilson \& Vallee}{1982}]{WV82} Wilson A., Vallee J., 1982, A\&AS, 47, 601
\bibitem[\protect\citeauthoryear{Windhorst et al.}{1984}]{Windhorst84} Windhorst R., va Heerde G., Katgert P., 1984, A\&A, 58, 1W
\bibitem[\protect\citeauthoryear{Windhorst et al.}{1985}]{Windhorst85} Windhorst R., Miley G., Owen F., Kron R., Koo D., 1985, ApJ, 289, 494
\bibitem[\protect\citeauthoryear{Windhorst et al.}{1990}]{Windhorst90} Windhorst R., Mathis D., Neuschaefer L., 1990, Astron. Soc. Pacific, 10, 398
\bibitem[\protect\citeauthoryear{Windhorst et al.}{1993}]{Windhorst93} Windhorst R., Fomalont E., Partridge R., Lowenthal J., 1993, ApJ,  405,  9498
\bibitem[\protect\citeauthoryear{Wise \& Abel}{2005}]{WA05} Wise J., Abel T., 2005, ApJ, 629, 615
\end{thebibliography}
\end{document}